\documentclass[journal]{IEEEtran}
\usepackage{graphicx}                                
\usepackage{amsmath}
\usepackage{multirow}
\usepackage[ruled,linesnumbered]{algorithm2e}
\usepackage{algorithmic}
\usepackage{cite}
\usepackage{amsthm}
\usepackage{amssymb}
\usepackage{array}
\usepackage{setspace}
\usepackage{booktabs}
\usepackage{float}
\usepackage{threeparttable}
\usepackage{caption,subcaption}
\usepackage{hyperref}
\usepackage{diagbox}
\usepackage{amsmath} 
\usepackage{multirow}
\usepackage{pifont}
\usepackage[normalem]{ulem}
\useunder{\uline}{\ul}{}
\usepackage[table,xcdraw]{xcolor}

\hypersetup{hidelinks,
	colorlinks=true,
	allcolors=black,
	pdfstartview=Fit,
	breaklinks=true}

\begin{document}
%
\title{End-to-end Hyperspectral Image Change Detection Network Based on Band Selection}
%
%
%
\author{
        Qingren Yao,
        Yuan Zhou,~\IEEEmembership{Senior Member,~IEEE,}
        Chang Tang,~\IEEEmembership{Senior Member,~IEEE,}
        and Wei Xiang,~\IEEEmembership{Senior Member,~IEEE}

\thanks{(\emph{Corresponding author: Yuan Zhou.})}

\thanks{Qingren Yao and Yuan Zhou are with the School of Electrical and Information Engineering, Tianjin University, Tianjin 300072, China (e-mail: qingren@tju.edu.cn; zhouyuan@tju.edu.cn).}
\thanks{Chang Tang is with the School of Computer
Science, China University of Geosciences, Wuhan 430079, China (e-mail: tangchang@cug.edu.cn).}
\thanks{Wei Xiang is with the School of Computing, Engineering and Mathematical Sciences, La Trobe University, Melbourne, VIC 3086, Australia, and also with the College of Science and Engineering, James Cook University, Cairns, QLD 4878, Australia (e-mail: w.xiang@latrobe.edu.au).}

}

\markboth{IEEE Transactions on Geoscience and Remote Sensing,Vol.XX, No.XX, XXXX}%
{Shell \MakeLowercase{\textit{et al.}}: Bare Demo of IEEEtran.cls for IEEE Journals}

\maketitle
\begin{abstract}
      For hyperspectral image change detection (HSI-CD), one key challenge is to reduce band redundancy, as only a few bands are crucial for change detection while other bands may be adverse to it. However, most existing HSI-CD methods directly extract change feature from full-dimensional HSIs, suffering from a degradation of feature discrimination. To address this issue, we propose an end-to-end hyperspectral image change detection network with band selection (ECDBS), which effectively retains the critical bands to promote change detection. The main ingredients of the network are a deep learning based band selection module and cascading band-specific spatial attention (BSA) blocks. The band selection module can be seamlessly integrated with subsequent CD models for joint optimization and end-to-end reasoning, rather than as a step separate from change detection. The BSA block extracts features from each band using a tailored strategy. Unlike the typically used feature extraction strategy that uniformly processes all bands, the BSA blocks considers the differences in feature distributions among widely spaced bands, thereupon extracting more sufficient change feature. Experimental evaluations conducted on three widely used HSI-CD datasets demonstrate the effectiveness and superiority of our proposed method over other state-of-the-art techniques.
\end{abstract}


\IEEEpeerreviewmaketitle

\section{Introduction}

\IEEEPARstart{H}{yperspectral} images (HSIs) are increasingly widely used due to their high spectral and spatial resolution. They typically comprise hundreds of continuous spectral bands in the ultraviolet, visible, near-infrared, and mid-infrared regions of the electromagnetic spectrum. Compared to other remote sensing images, such as visible images, multispectral images, and synthetic aperture radar images, HSIs can capture richer spectral information about ground objects. HSI change detection (CD) aims to recognize differences in multitemporal HSIs taken at the same geographic location. In recent years, HSI-CD has received increasing attention and found broad applications in areas such as forestry and agricultural monitoring \cite{gandhi2015ndvi}, natural disaster detection \cite{washaya2018coherence}, and land surface dynamic analysis \cite{yan2019time}.

HSI-CD algorithms can be classified into three categories as follows: algebra based, machine learning based and deep learning based algorithms. 
The algebra based methods directly calculate the difference between image pixels, and then determine the changed pixels through selecting an appropriate threshold \cite{du2012fusion, zhuang2016strategies}. However, these methods have high requirements for image preprocessing, and it is difficult to find appropriate thresholds. These deficiencies limit the prevailing of algebra based CD methods. 
Machine learning based approaches use hand-crafted feature extraction methods to convert HSIs into feature space, then contrast the features from bitemporal images \cite{wu2012hyperspectral, celik2009unsupervised}. But it’s difficult to effectively extract discriminative information through hand-crafted extraction process, resulting in an insufficient performance.
In contrast to machine learning, deep learning can extract features more effectively from complex high-level change information. Common neural networks such as convolutional neural networks (CNNs), recurrent neural networks (RNNs), graph convolutional networks (GCNs), etc. have been widely used in HSI-CD. The methods mostly focus on tuning network architecture to improve change feature extraction from full-dimensional HSIs. 

However, HSIs inherently contain redundant bands that are disadvantageous to change detection. Thereby, directly extracting features from the full-dimensional HSI can lead to feature confusion and reduce the discrimination of change feature. 

To address this issue, we propose an end-to-end hyperspectral image change detection network based on band selection (ECDBS), where only the bands that are favorable to HSI-CD are retained for improving change detection performance. ECDBS mainly consists of a novel deep learning based band selection module and cascading band-specific spatial attention (BSA) blocks. The band selection module is seamlessly integrated with a subsequent CD model, allowing joint optimization and end-to-end inference, rather than as a pre-processing step separate from change detection. It learns to measure the importance of each band based on the internal correlations between bands. The final low-dimensional HSI is obtained via clustering all bands and selecting the most important band for each cluster. The BSA block can adopt a tailored strategy to extract features of each band, taking into account the differences in feature distributions of widely spaced bands. It can generate a spatial attention map for each band to sufficiently capture change information, improving the discrimination of feature. The main contributions of this paper are summarized as follows:

\begin{enumerate}
        \item We design an end-to-end hyperspectral image change detection network based on band selection (ECDBS), aiming to mitigate band redundancy thereby improving HSI-CD performance. The network integrates a novel deep learning based band selection module. It can selectively retain the bands containing sufficient change information, facilitating subsequent change feature extraction.
        
	\item We streamline the pipeline of change detection incorporating band selection. The proposed HSI-CD network allows end-to-end inference as well as joint optimization of band selection and change detection, eliminating the need for a separate band selection step.

        \item To enhance change feature extraction, we propose a band-specific spatial attention (BSA) block that generates tailored spatial attention map for each band. Different from the widely used uniform feature extraction, the block can extract the features from each band with a strategy tailored to its feature distribution.

        \item Comprehensive experiments are performed on three widely adopted HSI-CD datasets. The experimental results demonstrate the superiority of our proposed model over other state-of-the-art HSI-CD methods.
\end{enumerate}

\section{Related Work}

Currently, there have been many HSI-CD algorithms. Generally, they can be categorized into the following three categories, i.e., algebra based, machine learning based, and deep learning based methods. In this section, we briefly describe the three sorts of algorithms in the previous works.

\subsection{Algebra based Method}

The earliest algebra based CD methods mainly include image difference method, image ratio method, and image regression method. When applying them to HSI-CD, choosing the optimal band is always a problem. Change vector analysis (CVA) \cite{malila1980change} is the most representative algebra based method. It firstly generates a difference image by calculating the Euclidean distance between the pixels of two images. Subsequently, the pixels can be classified by comparing the difference and a threshold. But CVA ignores the similarity between adjacent pixels. To overcome this deficiency, Thonfeld et al. \cite{thonfeld2016robust} proposed robust CVA (RCVA) which considers the influence of neighborhood pixels. Besides, there are several other algebra based methods \cite{kwan2019methods}, such as absolute average difference, vector for angle, and normalized Euclidean distance. These methods typically require a threshold to determine changed and unchanged pixels. The appropriate threshold could be determined manually or by using automatic threshold search methods such as expectation maximization \cite{bruzzone2000automatic} or K-means \cite{chen2017spectrally}, etc. Overall, algebra based methods are easy to implement and valid for easy scene of HSIs. But the accuracy of the radiation and geometric calibration has an important impact on the results of the algebra based methods.

\subsection{Machine Learning based Method}

Machine learning based methods project the HSIs into a new feature space, where the changed features are highlighted and unchanged features are suppressed. The widely used methods include principal component analysis (PCA) \cite{ortiz2006change}, multivariate alteration detection (MAD) \cite{nielsen1997multivariate}, and slow feature analysis (SFA) \cite{wu2014slow}. PCA can map data to the direction with the largest variance. PCA with k-means (PCAKM) \cite{celik2009unsupervised} uses PCA to generate low-dimensional features and perform k-means clustering on the reduced features to obtain change detection results. MAD uses canonical correlation analysis \cite{hardoon2004canonical} to maximize the correlation between the features of multitemporal images. The change detection results can be obtained by postprocessing the maximum correlation factor. Nielsen \cite{nielsen2007regularized} extended MAD to an iteratively reweighted MAD (IR-MAD), which conducts the weighted iteration according to the chi-square distance. SFA extracts the most temporally invariant component from the bitemporal images to transform the data into a new feature space. Multilayer cascade screening strategy ($\rm MCS^{4}CD$) \cite{liu2022multilayer} selects the highly reliable unchanged pixels as training samples and extracts the change information by iterative SFA. Besides, some methods utilize machine learning based classifier to achieve change detection. One type is direct classification method, and the other is post classification method. The direct classification method is to combine the bitemporal images together, and then a classifier, such as K-nearest neighbors (KNNs) \cite{abeywickrama2016k} or support vector machine (SVM) \cite{cortes1995support}, is used to find the changing categories. The post classification method first learns and classifies the bitemporary images, respectively, and then compares and analyzes the changes \cite{ahlqvist2008extending}. Although such kind of methods achieve effective dimensionality reduction and noise decline, they need consume an excessive number of computations when deal with a vast area.

\begin{figure*}[!t]
\centering
\includegraphics[width=\linewidth]{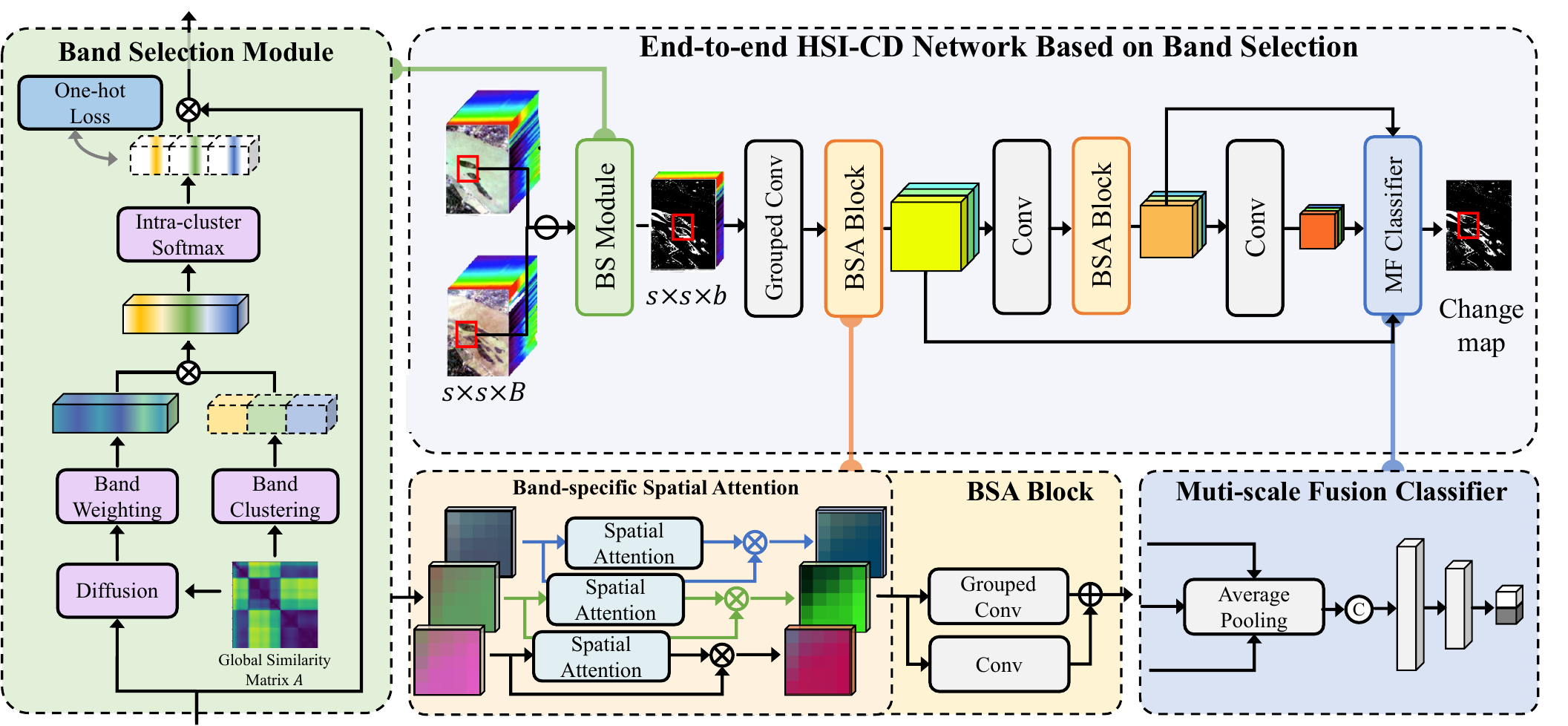}
\caption{Architecture of the proposed end-to-end hyperspectral image change detection network based on band selection (ECDBS).}
\label{fig1}
\end{figure*}

\subsection{Deep Learning based Method}

Deep learning based method is currently the mostly used HSI-CD method, as its strong capability of feature learning. A lot of common neural networks have been applied to this field, such as convolutional neural networks (CNNs), recurrent neural networks (RNNs), graph neural networks (GCN), and transformer. 

CNNs can fully extract the spatial-spectral features of images; therefore, it is widely used in HSI-CD. Wang et al. \cite{9494085} proposed SSA-SiamNet that integrates channel and spatial attention in CNN to highlight influential information and suppress less informative channels and pixels in the spectral and spatial domains. Gong et al. \cite{9664536} proposed a spectral and spatial network, which is capable to suppress CD-irrelevant spectral and spatial information via adaptive spectral and spatial mechanisms. Wang et al. \cite{wang2022rscnet} designed a residual self-calibrated network that can effectively exploit spatial information.  Yang et al. \cite{yang2022deep} proposed a deep multiscale pyramid network enhanced with spatial-spectral residual attention that aggregates low- and high-level features to enrich spatial details and semantic information. Luo et al. \cite{luo2023multiscale} fused spatial features from different scales with attention mechanism to generate a more discriminative change feature. To tackle the band redundancy, Zhan et al. \cite{zhan2021sscnn} proposed a spectral–spatial convolution neural network with a Siamese architecture (SSCNN-S) for HSI-CD, which employs 1D convolution to extract the spectral features. Li et al. \cite{li2023cbanet} proposed to preserve spectral characteristics from high data dimensionality through cross-band feature extraction and 2D self-attention. Ou et al \cite{ou2022cnn} introduced band selection to change detection, achieving remarkable performance. It proceeds in two stages, first selecting a band subset based on slow feature analysis and then constructing a difference matrix for change detection.

The RNNs that use cyclic connections for each neuron activation at each time step, showing great potential in processing continuous multi-temporal data. Qu et al. \cite{qu2021multilevel} integrated the architecture of encoder-decoder and long short-term memory network (LSTM) to simultaneously learn spatial-spectral features and temporal features. Shi et al. \cite{shi2022learning} devised a multi-path convolutional LSTM that enables the model can learn multi-level temporal dependencies of bitemporal HSIs. 

Recently, GCNs which extend convolution to irregular data structures, have gradually become a research hot spot in the field of HSI-CD. Qu et al. \cite{qu2021dual} proposed a dual-branch difference amplification graph convolutional network, which firstly introduced GCN to HSI-CD. Wang et al. \cite{wang2022csdbf} designed a dual-branch framework (CSDBF) that used CNN to extract pixel-level features and graph attention network to process superpixel-level features. 

Moreover, transformer-based HSI-CD models also attracted huge attention due to the advantage of modeling the long-range dependencies. Wang et al. \cite{wang2022spectral} designed a dual-branch transformer for HSI-CD that contains a spectral-spatial transformer and a temporal transformer (SST-Former) and evaluated the model is still valid in small training datasets. Wang et al. \cite{wang2023tritf} devised a triplet transformer framework based on parents-temporal and bother-spatial attention for HSI-CD. Although deep learning based algorithms have shown significant performance advantages, they still mostly rely on full-dimensional HSIs. This not only limits the potential performance of deep learning models but also reduces the efficiency of training and inference. 

\section{Methodology}

Given the dual-temporal hyperspectral remote sensing images, we subtract one from the other to obtain the difference image $I\in \mathbb{R}^{B\times h\times w}$, where $h$, $w$ represent the height and width of a HSI, and $B$ denotes the number of bands. Then we divide the entire difference HSI $I$ into patches as inputs of ECDBS, denoted $X\in \mathbb{R}^{B\times s\times s}$, where $s$ refers to the length and width of a patch.

As shown in Fig. \ref{fig1}, the ECDBS encompasses a band selection (BS) module, cascading band-specific spatial attention (BSA) blocks, and a multi-scale fusion classifier. The band selection module could select a few bands that contribute to change detection from all bands, producing a low-dimensional HSI. Subsequently, we extract the spatial feature from low-dimensional HSI through cascading BSA blocks. The blocks could enhance the feature discrimination through adopting the feature extraction strategy tailored to each band. Finally, the features with different scales are integrated and fed into fully connected layers to produce a change detection result in the multi-scale fusion classifier. In this section, the three main parts of the proposed ECDBS and the well-designed loss, are introduced in the following.

\subsection{Band Selection Module}
The high-dimensional and redundant spectral bands of HSI provide a large amount of information for change detection, but inevitably increase the difficulty of mining changed features. To address this issue, we propose the deep learning based band selection module. It can learn to measure band importance based on the internal correlations between bands. Considering the amount of information is also important for the selected band subset, we introduce band clustering in the band selection to avoid the highly similar bands are selected, i.e., grouping all bands into a number of clusters and then selecting the most important band from each cluster according to the learned band importance. The selection process contains three steps: firstly, we adopt a diffusion operation to shrink the difference among patches’ band correlation, promoting the learning of consistent band importance weight. Subsequently, the band weighting operation utilizes the band correlation to produce the band importance weights, indicating the amount of each band’s contribution to change detection. Finally, we apply SoftMax on the band importance weights in each cluster separately. The band with the highest importance weight in each cluster would be selected to form the final band subset. The detailed introduction to the above steps is given as follows.

\textbf{Band clustering} can assign the similar bands into a cluster and relatively different bands into different clusters. Selecting one band from each cluster allows the reserved information is as comprehensive as possible. Therefore, we introduce the band clustering in the band selection module to ensure that the selected band subset is still informative while the number of bands is reduced. To this end, we first construct a $B \times B$ similarity matrix $A$ based on a widely used strategy \cite{sui2021unsupervised}, where the similarity between bands is defined as follows:

\begin{equation} \label{eq1}
    A_{ij} = \left \{
    \begin{array}{ll}
        \frac{e_{i,k+1} - e_{ij}}{\sum^{k}_{m=1}(e_{i,k+1} - e_{i,m})}, &I_j \in \mathcal{N}_k(I_i) \\
        0, & \rm otherwise 
    \end{array} \right.
\end{equation}

\noindent where $e_{i,k+1}$ represents the $(k + 1)$th smallest Euclidean distance between the $i$th band $I_i$ and the other bands in $I$. The Euclidean distance between $i$th band and $j$th one is denoted as $e_{ij}$. $\mathcal{N}_{k}(I_{i})$ refers to the k-nearest neighbors of the band $I_{i}$. Using the similarity matrix $A$, we perform spectral clustering to partition all bands into $b$ clusters. The clustering result is recorded through an assignment matrix $C\in \mathbb{R}^{b\times B}$, where the row of the assignment matrix represents the indexes of the clusters, and the column represents the indexes of the bands. Formally, the assignment matrix is expressed as follows:

\begin{equation} \label{eq2}
C_{ij}=\begin{cases} 1, \quad I_{j}\in c_{i}\\ 
0, \quad \textit{otherwise}\end{cases}
\end{equation}

\noindent where $I_{j}$ denotes the $j$th band, and $c_{i}$ represent the $i$th cluster.

\textbf{Diffusion}. Influenced by the differences among patches, it’s challenging to learn a consistent band importance weight. Therefore, we adopt a diffusion operation to bring the band correlation of each patch closer. It can be expressed as follows :

\begin{equation}\label{eq3}
\overline{X}=\hat{A}XW
\end{equation}

\noindent where $\hat{A}$ is normalized similarity matrix from $A$, and $W$ is the learnable parameter. In this diffusion operation, each band absorbs the features from other bands according to the similarity defined by the matrix $\hat{A}$. Thus, the band correlation of all patches will approximate the band correlation of the entire HSI, promoting the learning of consistent band importance weight.

\textbf{Band weighting} exploits band correlation information to produce a band importance weight for each band. The importance weight indicates band's contribution to change detection. Specifically, we firstly construct a similarity vector $\boldsymbol{s} = \{ \boldsymbol{s}_{i \dots B}\} \in \mathbb{R} ^ {B}$ to represent the band correlation as follows:

\begin{equation}\label{eq4}
\boldsymbol{s}_{i}={\sum }_{i,j=1}^{B}\left| \overline{X}_{i}-\overline{X}_{j}\right|    
\end{equation}

\noindent where Euclidean distances is taken as the similarity metric and $\boldsymbol{s}_i$ is the sum of the Euclidean distances between the $i$th band and the others. To prevent the biased magnitude of similarity between different samples, we need to normalize $\boldsymbol{s}$ over a patch:

\begin{equation}\label{eq5}
\hat{\boldsymbol{s}}_{i} =\frac{\boldsymbol{s}_{i}-\mu _{\boldsymbol{s}}}{\sigma _{\boldsymbol{s}}+\epsilon }*\gamma +\beta  
\end{equation}

\noindent where $\hat{\boldsymbol{s}}_{i}$ is the normalized similarity between one band and all other bands, $\mu _{\boldsymbol{s}}$ is the expected value, $\sigma _{\boldsymbol{s}}$ is the standard deviation, $\epsilon $ (e.g., 1e-5) is a constant added for numerical stability, and $\gamma $ and $\beta $ are learnable affine transform parameters. Subsequently, the normalized similarity vector $\hat{\boldsymbol{s}}=\left\{\hat{\boldsymbol{s}}_{i\ldots B}\right\}\in \mathbb{R}^{B}$ is fed into two fully connected layers to guide the learning of band importance weights $\boldsymbol{w} \in \mathbb{R}^{B}$. It can be formulated as follows:

\begin{equation}\label{eq6}
\boldsymbol{w}= {\rm Sigmoid} (W_{1} {\rm ReLU} (W_{0}\hat{\boldsymbol{s}}))
\end{equation}

\noindent where $W_{0}$, $W_{1}$ are the learnable parameters.

\textbf{Intra-cluster SoftMax}. To select one band from each cluster, we multiply the band weight $\boldsymbol{w}$ with the assignment matrix $C$ row by row to obtain a weight matrix $\mathcal{W}\in \mathbb{R}^{b\times B}$. Afterward, we perform Softmax on the non-zero elements of each row to highlight one band that contributes most to change detection in each cluster, resulting in a selection matrix $E\in \mathbb{R}^{b\times B}$, which is defined as:

\begin{equation}\label{eq7}
E_{ij}=\frac{\exp (\mathcal{W}_{ij}/\tau )}{\sum _{k}\exp (\mathcal{W}_{ik}/\tau )}, \mathcal{W}_{ij}\neq 0, \mathcal{W}_{ik}\neq 0
\end{equation}

\noindent where $\tau $ is the temperature parameter that decreases with epoch. As the temperature parameter becomes smaller, each row vector of selection matrix becomes sparser and sparser, eventually approximating a one-hot vector.

Finally, using the selection matrix, we select the band with the largest importance weight from each cluster.  The formulaic representation is as follows:

\begin{equation}\label{eq8}
\chi =E\times X
\end{equation}
\noindent where $\chi \in R^{b\times s\times s}$ is the patch after dimensionality reduction. 

\begin{figure*}[t]
\centering
\includegraphics[width=\linewidth]{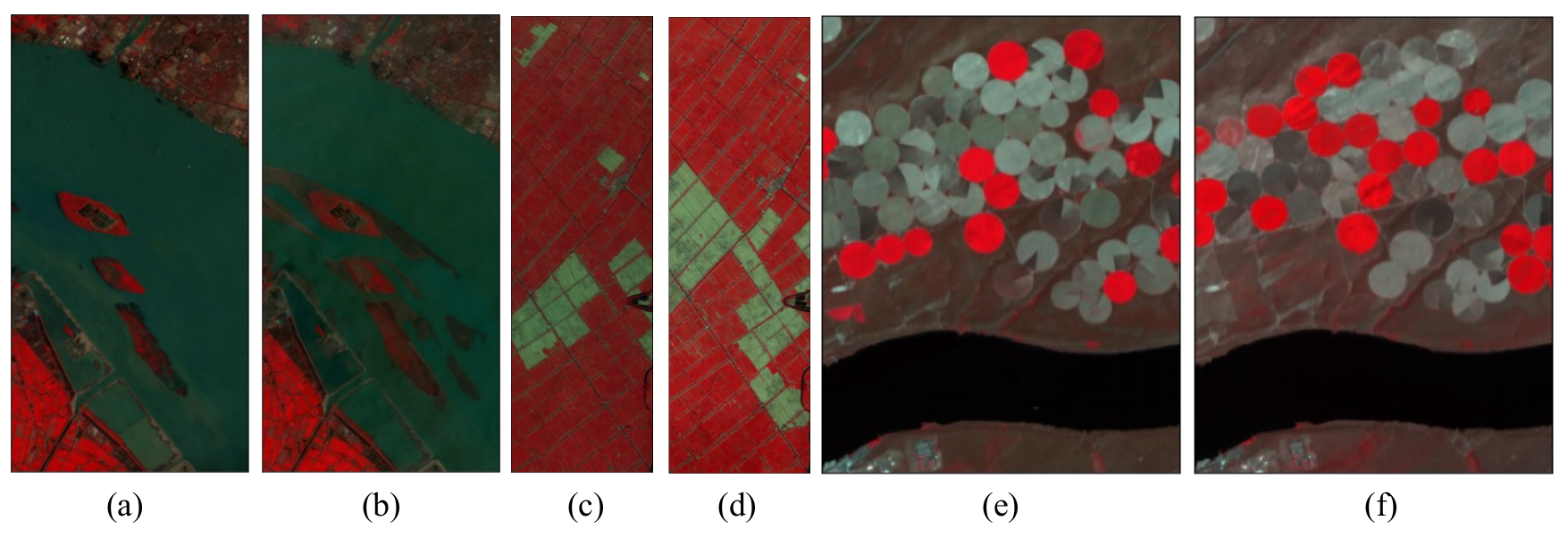}
\caption{Three datasets used in experiments: (a) and (b) are River images acquired on May 3, 2013 and December 31, 2013; (c) and (d) are Farmland images acquired on May 3, 2006 and April 23, 2007; (e) and (f) are USA images on May 1, 2004 and May 8, 2007.}
\label{fig_datasets}
\end{figure*}

\subsection{Band-specific Spatial Attention}

During feature extraction, most CD models uniformly process all bands, ignoring the differences in feature distribution of different bands. This would lead to a degradation of feature discrimination when processing HSIs after band selection, as the feature distributions of different bands after band selection are significantly different. To address this issue, we design a band-specific spatial attention mechanism that generates tailored spatial attention map for each band based on its feature distribution. Specifically, within each band, we determine the importance of a spatial position by measuring the similarity between its feature and the global spatial feature of the band.

Given the patch feature $\chi = \{ \chi_i \}$, $i\in[1:b]$ where $b$ bands are from different clusters. Each band have a vector representation at every position, namely  $\chi _{i}=\left\{\boldsymbol{x}_{1\ldots m}\right\}$, where $m=s \times s$, $\boldsymbol{x}_{j}$ is the spatial vector at position $j$. For a certain band, we firstly construct the global spatial vector $\boldsymbol{g}$ through spatial average pooling:

\begin{equation}\label{eq9}
\boldsymbol{g}=\frac{1}{m}{\sum }_{j=1}^{m}\boldsymbol{x}_{j}
\end{equation}

\noindent With the global spatial vector, the importance coefficient of each position could be calculated. This is achieved by dot product, which measures the similarity between the global spatial vector $\boldsymbol{g}$ and the local spatial vector $\boldsymbol{x}_{j}$ to some extent. Thereby for each position, we have:

\begin{equation}\label{eq10}
c_{j}=\boldsymbol{g}\cdot \boldsymbol{x}_{j}
\end{equation}

\noindent To avoid the magnitude bias among various patches, we normalize the importance coefficient $c_j$ in the way as same as Eq. (\ref{eq5}), resulting in a normalized importance coefficient $a_j$. Finally, the original $\boldsymbol{x}_{j}$ is scaled by the generated importance coefficient $a_{j}$ via a Sigmoid function over the space:

\begin{equation}\label{eq12}
\hat{\boldsymbol{x}}_{j}=\boldsymbol{x}_{j}\cdot {\rm Sigmoid} \left(a_{j}\right)
\end{equation}

\noindent where $\hat{\boldsymbol{x}}_j$ is the enhanced feature at position $j$. The enhanced features at all locations collectively constitute the enhanced feature of a band $\hat{\chi}_i$. And the resulted patch feature is composed of the enhanced features from all bands, i.e., $\hat{\chi} = \{\hat{\chi}_i\}$.

As the superior performance of ResNet \cite{he2016deep}, we embed the band-specific spatial attention and grouped convolution into a residual block, which we refer to as the BSA Block, as illustrated in Fig. \ref{fig1}. Following each BSA Block, a convolution without padding is applied to increase the receptive field and decrease the output feature map size, which has been widely used in handling patch data with a small size \cite{9494085}.

\begin{figure*}[t]
\centering
\includegraphics[width=\linewidth]{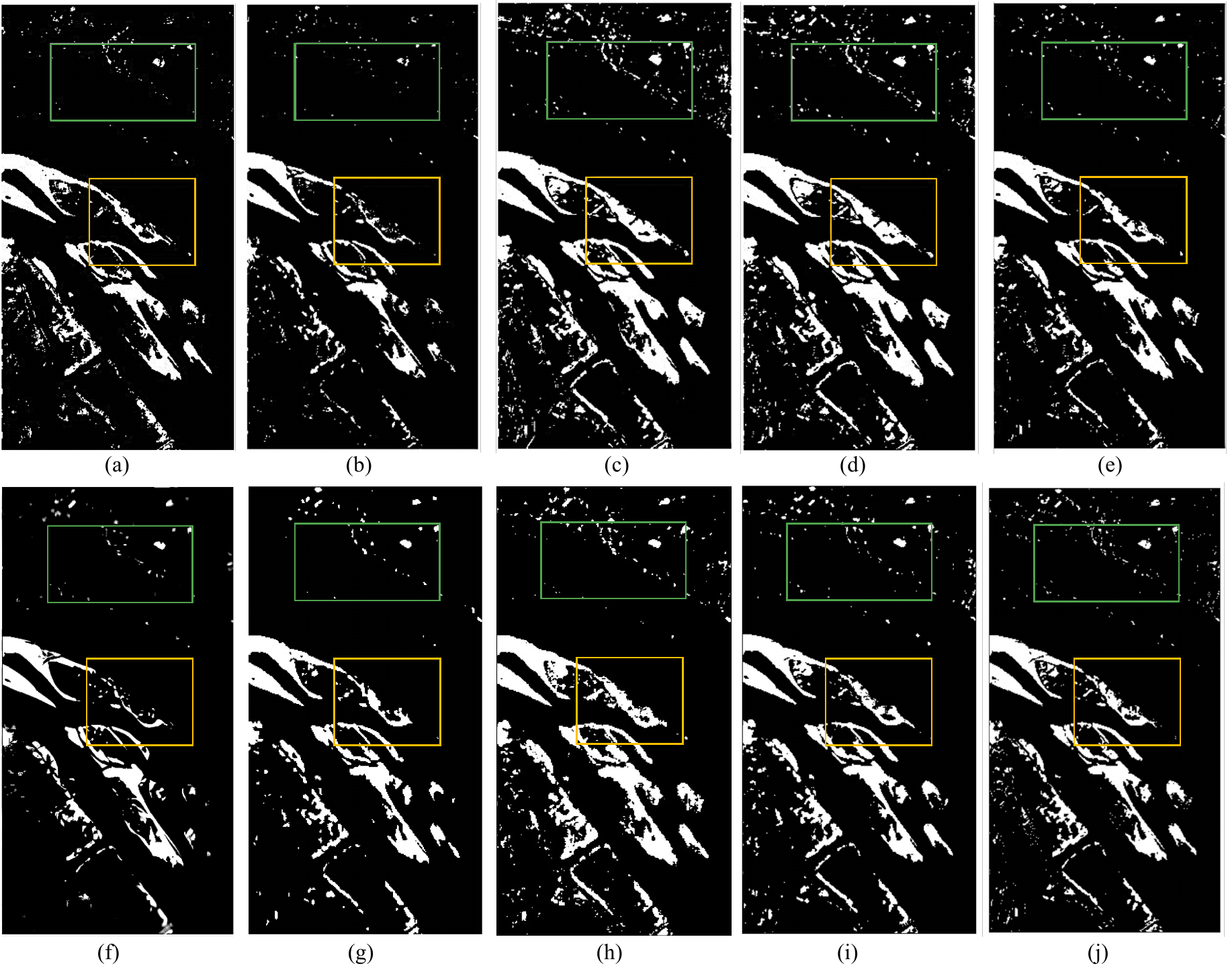}
\caption{ Change detection maps for the River dataset. (a) PBCNN (b) 3D-CNN (c) GETNET (d) SSCNN (e) SSA-SiamNet (f) SFBS-FFGNET (g) CSDBF (h) CBANet (i) Ours (j) Ground truth.}
\label{fig_rivercdmap}
\end{figure*}

\subsection{Multi-scale Fusion Classifier}

In change detection or visual recognition, integrating features with different scales has been proved to be an effective solution to improve feature discriminativeness and parameter utilizing efficiency \cite{ou2022cnn, yang2022deep}. Therefore, we aggregate the features from different depths with different scales to form a comprehensive feature $\boldsymbol{x}_{a}$: 

\begin{equation}\label{eq13}
\boldsymbol{x}_{a}=\left[{\rm avg} \left(x_{b1}\right), {\rm avg} \left(x_{b2}\right), {\rm avg} \left(x_{b3}\right)\right]
\end{equation}

\noindent where $x_{b1}$, $x_{b2}$ and $x_{b3}$ are the features from the first, second BSA Block and the final convolution layer respectively, the $\rm avg()$ denotes the spatial average pooling, $[\cdot ]$ represents the concatenation operation. Then the feature is imported to the two fully connected layers to generate the final change detection result. Formally, the multi-scale fusion classifier can be expressed as:

\begin{equation}\label{eq14}
f\left(\boldsymbol{x}_{a}\right) = {\rm Softmax} \left( \Theta_{2}\left(\Theta_{1}\boldsymbol{x}_{a}+b_{1}\right)+b_{2}  \right) 
\end{equation}

\noindent where $f()$ denotes the two fully connected layers, $\Theta $ and $b$ are the learnable parameters of the fully connected layers.

\subsection{Loss Function}

The loss $L$ for optimizing our model consists of two items, a classification loss $L_{C}$ for supervising the change detection results and a selection loss $L_{E}$ used to used to constrain the selection matrix:

\begin{equation}\label{eq15}
L=L_{C}+\alpha L_{E}
\end{equation}

\noindent where $\alpha $ is a trade-off coefficient that balances the minimization between the classification error and constraint item.

We adopt the weighted binary cross entropy loss as the classification loss to handle the imbalance between the number of changed and unchanged samples, which is characterized as:

\begin{equation}\label{eq16}
L_{C}=-\left[\omega _{c}y\log \left(\hat{y}\right)+\omega _{u}\left(1-y\right)\log \left(1-\hat{y}\right)\right]
\end{equation}

\noindent where $\hat{y}$ represents the predicted probability, $y$ is the true label that is 1 for the changed sample and 0 for the unchanged sample, $\omega _{c}$ and $\omega _{u}$ are the weights of changed sample and unchanged sample, respectively.

The selection loss aims to guide the row vector of selection matrix $E$ approximates the one-hot vector so that the only one band is selected in each cluster. So, we regularize the entropy of the selection matrix by:

\begin{equation}\label{eq17}
L_{E}= - \frac{1}{b}{\sum }_{i}^{b}{\sum }_{j}^{B}E_{ij}\log (E_{ij})
\end{equation}
\noindent where $E_{ij}$ is the $i$-th row j-th column of selection matrix $E$.

\section{Experimental}

\subsection{Datasets}

Our method was evaluated on three widely used HSI datasets: Farmland, River, and USA. These datasets were collected using the Hyperion sensor onboard the EO-1 satellite. The sensor covers wavelengths ranging from 0.4 to 2.5 ${\mu}$m with 242 spectral bands, providing HSIs with a spatial resolution of approximately 30 m and a spectral resolution of approximately 10 nm. 

\subsubsection{River dataset} The River dataset covers a river region in Jiangsu Province. It includes bitemporal HSIs acquired on May 3, 2013, and December 31, 2013, with a spatial extent of 463 $\times$ 241 pixels. After removing noisy bands, 198 bands are selected for change detection. The bitemporal images are shown in Fig. \ref{fig_datasets}(a) and (b) (bands 33, 22, and 11 as RGB). Changes in the dataset are mainly caused by the removal of sediment from the river.

\subsubsection{Farmland dataset} The Farmland dataset consists of bitemporal scenes captured over a farmland in Yancheng, Jiangsu Province, China, on May 3, 2006, and April 23, 2007. The dataset has a spatial size of 450 $\times$ 140 pixels, and 155 bands were considered after removing spectral bands with low signal-to-noise ratios. The false-color composites (band 33, 22, and 11 as RGB) of the two images are shown in Fig. \ref{fig_datasets} (c) and (d). Changes in the dataset are mainly caused by crop rotation.

\subsubsection{USA dataset} The USA dataset was obtained from an irrigated agricultural field in Hermiston City, Umatilla County, OR, USA. The dataset includes two hyperspectral images taken in 2004 and 2007, respectively. This dataset has a spatial size of 307 $\times$ 241 pixels and contains 154 bands after removing noisy bands. The false-color composites (bands 33, 22, and 11 as RGB) of the bitemporal images are shown in Fig. \ref{fig_datasets}(e) and (f). Changes in the dataset are mainly caused by the regulation of irrigation areas.

\subsection{Experimental Setup}
\subsubsection{Sampling Rate and Compared Methods} Referring to the settings in \cite{wang2022csdbf, shi2022learning}, we randomly select  3.36\%, 20.95\%, and 9.77\% of pixels for training on the River, Farmland, and USA datasets, respectively. Approximately 1\% of these samples are reserved for validation, with the remaining data used for testing. To evaluate the effectiveness of the proposed method, we compare it against other commonly used and advanced approaches, including GETNET \cite{wang2018getnet}, PBCNN \cite{sharma2017patch}, 3D-CNN \cite{li2017spectral}, SSCNN \cite{zhan2021sscnn}, SSA-SiamNet \cite{9494085}, SFBS-FFGNET \cite{ou2022cnn}, CSDBF \cite{wang2022csdbf} and CBANet \cite{li2023cbanet}. 

\subsubsection{Evaluation metrics} To evaluate the performance of different methods, we utilize three commonly used quantitative assessment indices, including overall accuracy (OA), Kappa coefficient (Kappa), and F1-score (F1).

OA is the overall accuracy and the proportion of the correctly predicted pixels to the number of pixels in the whole image. The formula is written as follows:

\begin{equation}
\label{eq 181}
\rm OA = \rm \frac{TP+TN}{TP+TN+FP+FN}
\end{equation}

\noindent where True Positives (TP) denotes the number of correctly detected changed pixels, True Negatives (TN) represents the number of correctly identified unchanged pixels, False Positives (FP) refers to the number of false-alarm pixels, and False Negatives (FN) indicates the number of missed changed pixels.

Kappa coefficient is an accuracy measure of the classification task. The calculation of the coefficient is based on the confusion matrix, and the value is between $-1$ and 1. The closer the result is to one, the better the consistency of the classification. Kappa coefficient is calculated as follows:

{

\begin{equation} \label{eq 182}
\begin{aligned}
    & {\rm Kappa} = \frac{\mathrm{OA} - P_C}{1 - P_C}  \\
    & \! P_C = \rm \frac{\left(TP\! + \!FP\right)\left(TP\! + \! FN\right) + \left(FN\!+\!TN\right)\left(FP\!+\!TN\right)}{\left(TP+FP+TN+FN\right)^{2}}
\end{aligned}    
\end{equation}
}

F1 score is the harmonic mean of the precision and recall, so this criterion can balance the impact of precision and recall and reflect the average performance of the model.  When F1 is higher, it indicates that the CD method is more effective. The formula can be written as follows:

\begin{equation} \label{eq 183}
\begin{aligned}
    \rm F1 &= \rm 2\times \frac{Pr\times Re}{Pr+Re}  \\
    \rm Pr &= \rm \frac{TP}{TP+FP}  \\
    \rm Re &= \rm \frac{TP}{TP+FN}
\end{aligned}
\end{equation}

\begin{table}[t]
\centering
\caption{Change detection results of the river dataset, model parameters and FLOPs.}
\resizebox{\linewidth}{!}{
\begin{tabular}{lccccccc}
\hline
Method        & OA                      & Kappa                 & F1                        & Parameters(k)   & FLOPs(k)   \\ \hline
PBCNN(2017)         & 96.02                   & 71.61                 & 73.79                     & 235.17          & 3416.93   \\
3D-CNN(2017)        & 96.71                   & 73.76                 & 75.39                     & 245.79          & \textcolor{blue}{492.23}   \\
GETNET(2019)        & 95.51                   & 74.42                 & 76.76                     & 154179.42       & - \\
SSCNN-S(2021)       & 95.54                   & 74.08                 & 76.51                     & \textcolor{red}{21.68}        & 1082.72   \\
SSA-SiamNet(2022)   & \textcolor{blue}{97.35} & \textcolor{blue}{82.01} &\textcolor{blue}{83.36}  & 107.18       & 1795.66    \\
SFBS-FFGNET(2022)  & 96.70                 & 77.10                 & 82.09                     & 338.18       &  5445.22      \\
CSDBF(2022)         & 96.97                   & 79.82                 & 82.31                     & 122.41       & -     \\
CBANet(2023)    & 96.73 & 75.60 & 78.50 & 343.35 & 6791.19 \\
Ours & \textcolor{red}{97.46}  & \textcolor{red}{83.15} & \textcolor{red}{84.53}            & \textcolor{blue}{42.21}        & \textcolor{red}{288.64} \\ \hline
\end{tabular}}
\label{table1}
\end{table}

\begin{figure}[t]
\centering
\includegraphics[width=0.9\linewidth]{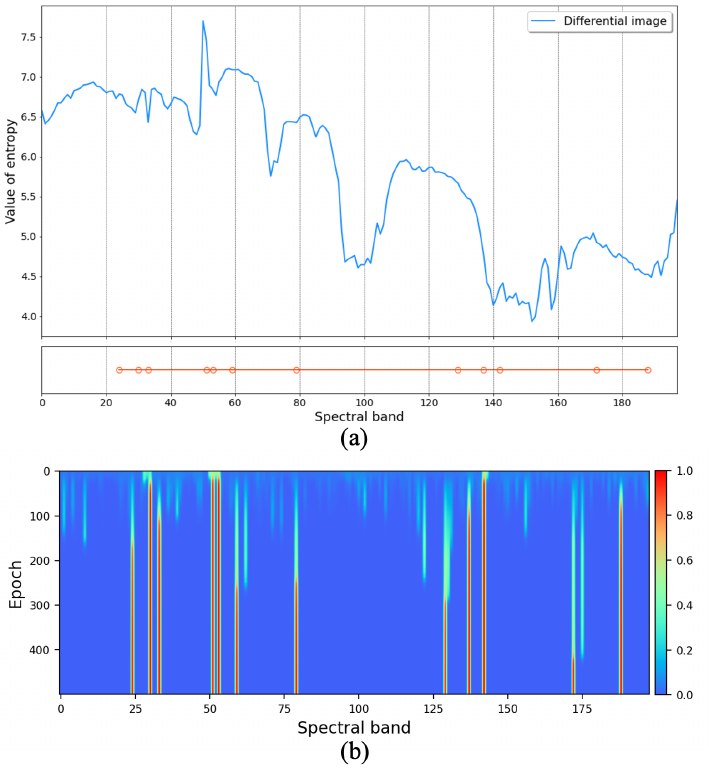}
\caption{(a) The 12 bands of River dataset selected by band selection module and the entropy value of each band in differential HSI. (b) Visualization of band importance weights under varying iterations in River dataset.}
\label{fig_rivercombo}
\end{figure}

\begin{figure*}[t]
\centering
\includegraphics[width=\linewidth]{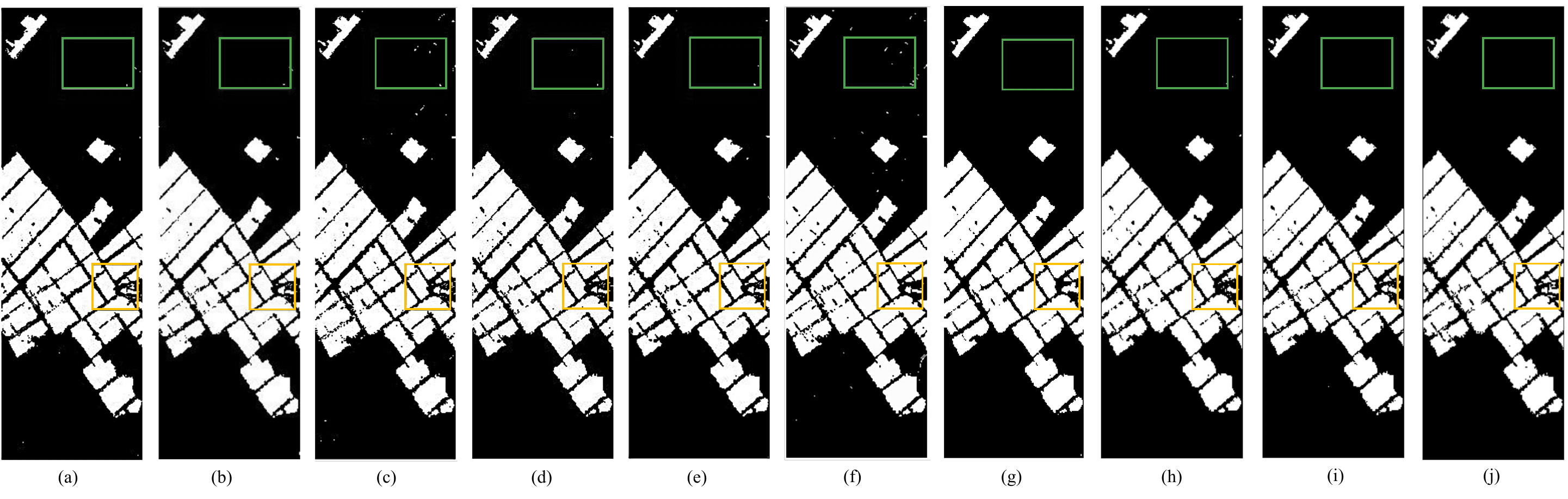}
\caption{ Change detection maps for the Farmland dataset. (a) PBCNN (b) 3D-CNN (c) GETNET (d) SSCNN (e) SSA-SiamNet (f) SFBS-FFGNET (g) CSDBF (h) CBANet (i) Ours (j) Ground truth.}
\label{fig_farmlandcdmap}
\end{figure*}

\subsubsection{Parameter Setting} In the proposed method, several hyperparameters, such as the number of selected bands, the number of channels in feature map, and the trade-off parameter $\alpha $ can affect the model's performance in the final change detection task. Hence, we analyzed the impact of these parameters in our hyperparameter analysis experiments, which is shown in IV-F. In comparison experiments, we set the band down sample rate as 16, i.e., selecting 12, 10, and 10 bands from the 198, 155, and 154 spectral bands in the River, Farmland, and USA datasets, respectively. The number of channels in feature map is set to three times the number of selected bands, i.e., 36, 30, and 30 for the River, Farmland, and USA datasets, respectively. For the trade-off parameter $\alpha $, we set it as 0.1 empirically. Considering the number of changed samples is much less than that of unchanged sample, we set the weights $\omega _{c}$ and $\omega _{u}$ of changed samples and unchanged sample to 5 and 1, respectively, to increase the penalty for changed samples. Furthermore, the hyperparameter settings of the comparable methods are kept consistent with those described in their respective articles. The input patch size is $5 \times 5$ for all methods except for GETNET and CSDBF which take the entire HSI as input. To avoid biased estimation, all experiments are  carried out using Pytorch in a single NVIDIA RTX 3060 GPU with 12-GB memory.

\begin{figure}[t]
\centering
\includegraphics[width=0.9\linewidth]{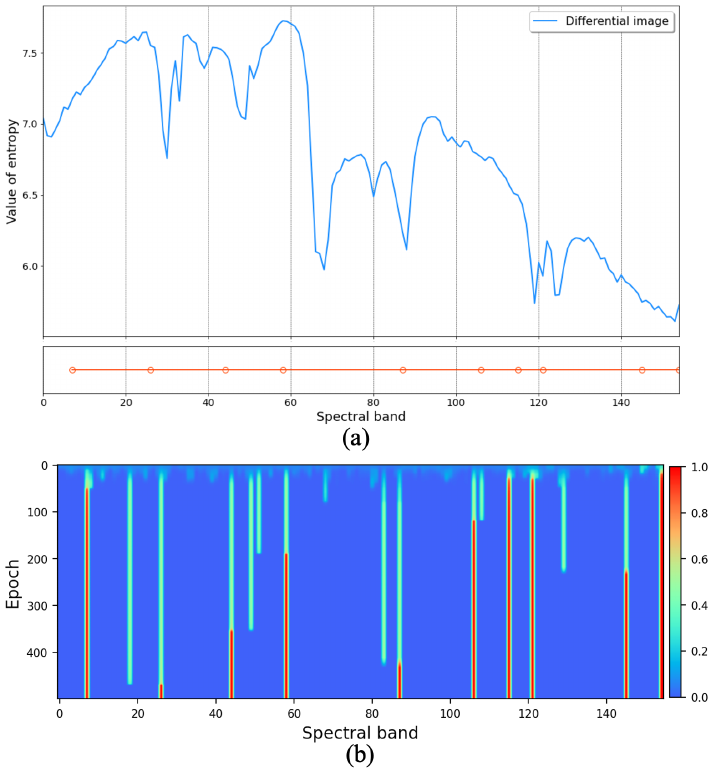}
\caption{(a) The 12 bands of Farmland dataset selected by band selection module and the entropy value of each band in differential HSI. (b) Visualization of band importance weights under varying iterations in Farmland dataset.}
\label{fig_farmlandcombo}
\end{figure}

\begin{figure*}[t]
\centering
\includegraphics[width=\linewidth]{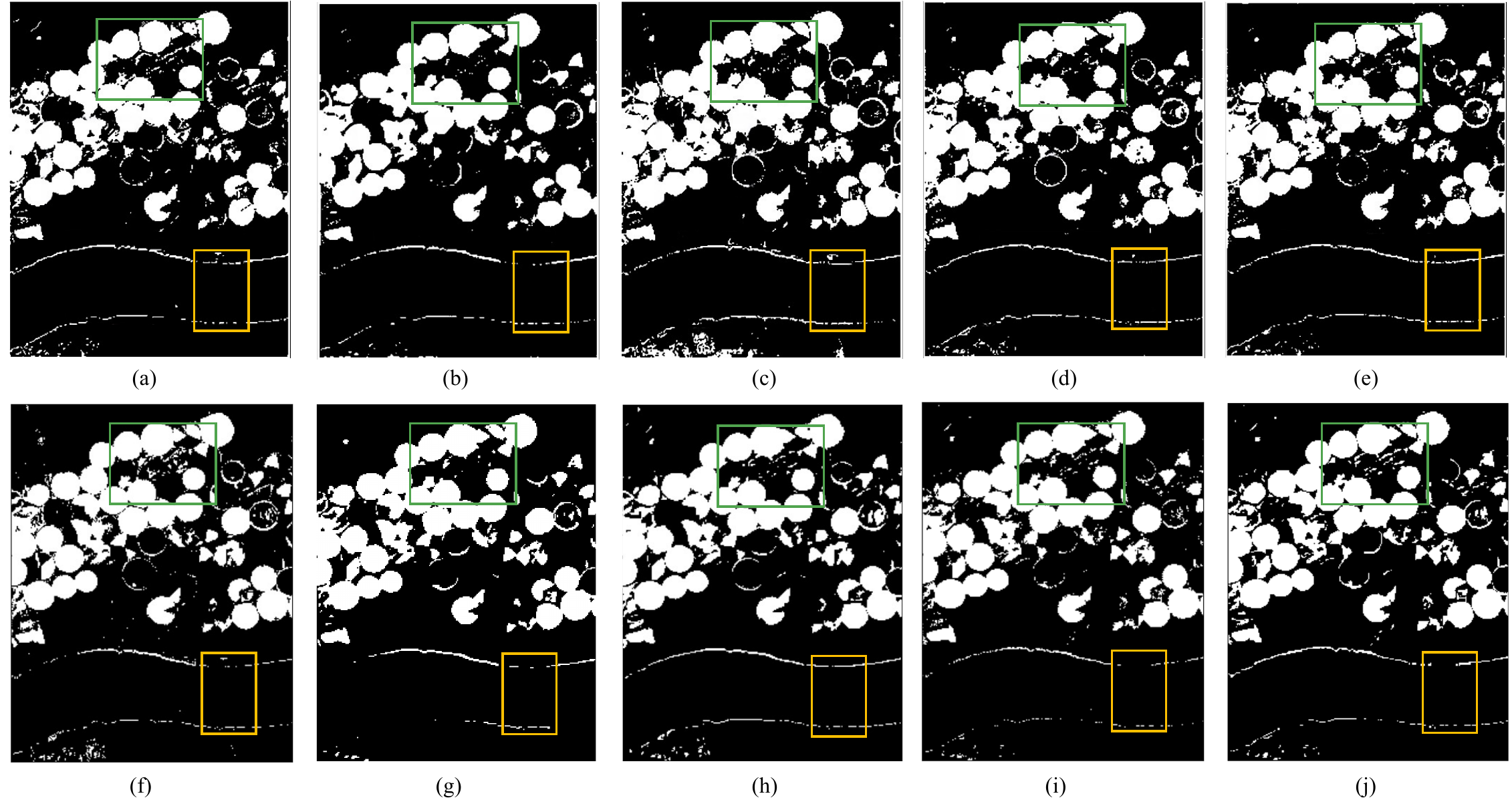}
\caption{ Change detection maps for the USA dataset. (a) PBCNN (b) 3D-CNN (c) GETNET (d) SSCNN (e) SSA-SiamNet (f) SFBS-FFGNET (g) CSDBF (h) CBANet (i) Ours (j) Ground truth.}
\label{fig_usacdmap}
\end{figure*}

\begin{table}[t]
\centering
\caption{Change detection results of the farmland dataset, model parameters and FLOPs.}
\resizebox{\linewidth}{!}{
\begin{tabular}{lccccccc}
\hline
Method           & OA               & Kappa & F1    & Parameters(k)  & FLOPs(k)   \\ \hline
PBCNN(2017)      & 97.61            & 94.13 & 95.79 & 210.40   & 2797.73   \\
3D-CNN(2017)      & 98.01           & 95.20  & 96.54 & 196.25   & \textcolor{blue}{393.16}    \\
GETNET(2019)      & 97.79           & 94.46 & 95.99 & 93415.26 & - \\
SSCNN-S(2021)     & \textcolor{blue}{98.45} & 96.13 & \textcolor{blue}{97.19} & \textcolor{red}{18.07}    & 902.12    \\
SSA-SiamNet(2022) & 98.12           & 95.81 & 96.34 & 88.60    & 1486.06   \\
SFBS-FFGNET(2022) & 97.60           & 94.30  & 95.94 & 338.18   & 3295.40         \\
CSDBF(2022)       & 98.43           & \textcolor{blue}{96.20}  & 97.06 & 105.64   & -      \\
CBANet(2023) & 98.03 & 95.11 & 96.47 & 332.34 & 6511.19 \\
Ours        & \textcolor{red}{98.58} & \textcolor{red}{96.47}  & \textcolor{red}{97.46} & \textcolor{blue}{30.68}    & \textcolor{red}{213.64}     \\ \hline

\end{tabular}}
\label{table2}
\end{table}

\subsection{Experiment on River Dataset}

Table \ref{table1} displays the results of testing all methods on the River dataset, where the proposed method achieves the best performance in terms of OA, Kappa and F1 indices. Due to class imbalance in the dataset, the OA values of most methods are close, but significant differences can be observed in Kappa values. The proposed method demonstrates a clear advantage in the Kappa index and achieved a 1.14\% improvement over the second-best method, SSA-SiamNet. Different from SFBS-FFGNET, which employs a band selection method that is totally independent with the subsequent change detection model, our model integrates a band selection module that is trained jointly with the CD model. It allows our model to select more discriminative bands for change detection, showing better performance. 

For evaluating the trade off between the complexity and accuracy of each model, the number of trainable parameters and the floating point of operations (FLOPs) of different models are also provided and listed in the last two columns of Table \ref{table1}. For fairness, we do not compare FLOPs with GETNET and CSDBF, because both methods inference with an entire image and typically cause higher FLOPs. Our proposed method shows fewer parameters and faster inference speed comparing with most patch-based methods. 

Fig. \ref{fig_rivercdmap} presents the change detection maps obtained by the different methods. For most methods, it can be observed that many small objects are misidentified over the green box and the change detection results in the yellow box do not match well with the ground truth. However, the change detection map obtained by our model is the closest to the ground-truth map.

The 12 bands selected by band selection module and the entropy of differential HSI are shown in Fig. \ref{fig_rivercombo} (a). We can see the 12 bands are carefully selected, avoiding adjacent bands containing high redundancy. It is attributed to clustering constraint that effectively partitions all bands into several groups with low correlation. The convergence process of band importance weight is illustrated in Fig. \ref{fig_rivercombo} (b). During the training process, those bands that are favorable to change detection are progressively highlighted, while bands that do not aid in change detection are suppressed. After 400 epochs, the 12 bands that are most beneficial to change detection in all bands are identified and the band importance weights become stable.

\subsection{Experiment on Farmland Dataset}

Table \ref{table2} presents the results of all experimental methods on the Farmland dataset. Our proposed method achieves the highest OA, Kappa, and F1, with increases of 0.13\%, 0.34\%, and 0.27\% respectively, over the second-best method, SSCNN-S. CSDBF also performs well, achieving the third-best performance with an OA of 98.43\%, a Kappa coefficient of 96.20\% and a F1 score of 97.06\%. Compared with SSA-SiamNet that employs one attention map to scale spatial features of all bands, our proposed model shows a significant improvement with a 0.46\% increase in OA and a 0.66\% increase in Kappa index. The possible reason is that our proposed model considers the differences in feature distributions among bands while feature extraction rather than uniformly treat all bands.

Analyzing the data in the last two columns of Table \ref{table2}, our proposed method uses fewer parameters to achieve better performance in change detection and possesses the highest inference speed. This superior performance can be attributed to our proposed band selection module and the band-specific spatial attention. The former eliminates spectral redundancy of HSIs so that our model does not need extra parameters to process the irrelevant bands. The latter improves the discrimination of changed features in the selected bands, enabling the accurate change detection based on low-dimensional HSIs. 

The change maps derived by different CD models are shown in Fig. \ref{fig_farmlandcdmap}. It can be observed that most methods display varying degrees of false detection for small objects in green box, but our model, CSDBF, and CBANet show more accurate classification in this regard. Furthermore, the change detection results of our method in the yellow box match better with the ground truth.

The distribution of selected 10 bands and the entropy of differential HSI are presented in Fig. \ref{fig_farmlandcombo} (a). As we can see, the selected bands not only distribute uniform but also avoid the most sharply decreasing regions with low entropy. And the convergence process of band importance weights is shown in Fig. \ref{fig_farmlandcombo} (b).

\subsection{Experiment on USA Dataset}

\begin{figure}[t]
\centering
\includegraphics[width=0.9\linewidth]{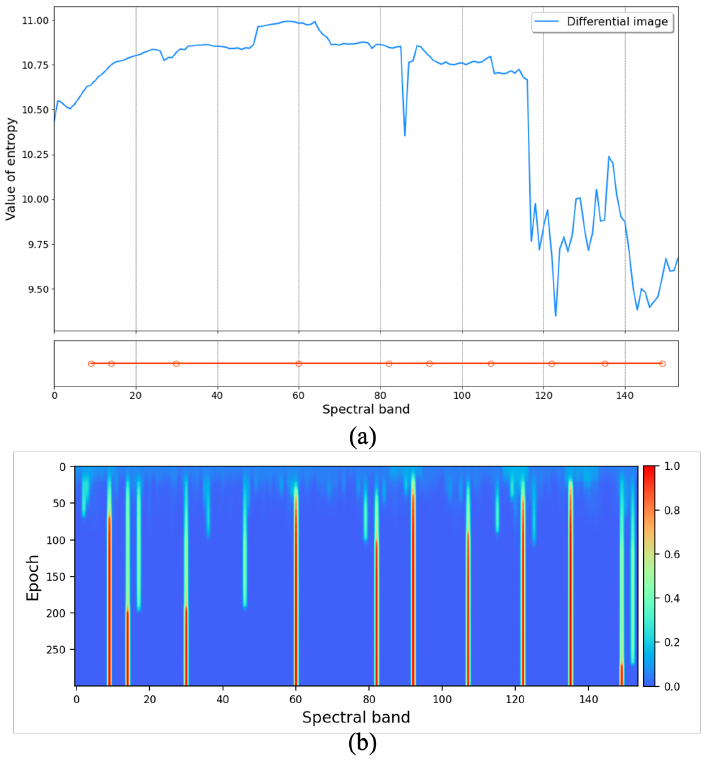}
\caption{(a) The 10 bands of USA dataset selected by band selection module and the entropy value of each band in differential HSI. (b) Visualization of band importance weights under varying iterations in USA dataset.}
\label{fig_USAcombo}
\end{figure}

Table \ref{table3} presents the change detection results of the different methods in the experiment on the USA dataset. Our method achieves the highest OA and F1, demonstrating better overall performance than other methods. Although CSDBF gives the highest Kappa, it is slightly inferior to our method in other metrics and contains five times the number of parameters as us. CBANet also demonstrates good performance, achieving the third best results with an OA of 96.64\%, a Kappa coefficient of 90.01\%, and a F1 score of 92.15\%. Moreover, we observe that SSCNN-S doesn’t achieve great performance with a lightweight network structure. The reason may be that it directly applies the $1\times1$ convolution layer to fuse the redundant bands to reduce the spectral dimension of HSI, making it difficult to obtain discriminative low-dimensional features in a limited sample range. We suppose that a deep learning based band selection module is a more sensible option to reduce the spectral dimension of HSI, because it can avoid the interfere from irrelevant bands and select discriminative bands. Fig. \ref{fig_usacdmap} presents the change maps obtained from different methods. As we can see, most methods fail to accurately detect the changed pixels in the center of green box. Moreover, a segment of the riverbank in the yellow box that has not changed is incorrectly detected as a change by most methods. Our method shows better performance in the two areas.
The distribution of selected 10 bands and the entropy of differential HSI are shown in Fig \ref{fig_USAcombo} (a). And the convergence process of band importance weights is presented in Fig \ref{fig_USAcombo} (b).

\begin{table}[t]
\centering
\caption{Change detection results of the USA dataset, model parameters and FLOPs.}
\resizebox{\linewidth}{!}{
\begin{tabular}{lccccccc}
\hline
Method      & OA    & Kappa & F1    & Parameters(k)  & FLOPs(k)   \\ \hline
PBCNN(2017)      & 95.72 & 86.80 & 89.43 & 209.83   & 2783.33   \\
3D-CNN(2017)      & 96.31 & 88.25 & 90.59 & 195.10     & \textcolor{blue}{390.86}   \\
GETNET(2019)      & 93.97 & 82.41 & 86.26 & 93415.26 & - \\
SSCNN-S(2021)     & 96.30 & 88.64 & 91.03 & \textcolor{red}{17.98}    & 897.92    \\
SSA-SiamNet(2022) & 95.80 & 87.41 & 90.01 & 88.22     & 1478.86   \\
SFBS-FFGNET(2022) & 95.99 & 88.25 & 90.82 & 338.18    & 3295.40       \\
CSDBF(2022)       & \textcolor{blue}{96.87} & \textcolor{red}{90.93} & \textcolor{blue}{92.70} & 105.25   & -     \\
CBANet(2023) & 96.64 & 90.01 & 92.15 & 332.08 & 6506.39 \\
Ours        & \textcolor{red}{96.95} & \textcolor{blue}{90.81} & \textcolor{red}{92.74} & \textcolor{blue}{29.19}     & \textcolor{red}{199.30}      \\ \hline
\end{tabular}}
\label{table3}
\end{table}

\subsection{Hyperparameter Analysis}

\begin{table}[t]
\centering
\caption{Change detection results of different down sample rates and expansion rates in three datasets.}
\resizebox{\linewidth}{!}{
\begin{tabular}{c|cc|ccc}
\hline
Dataset                   & Downsample & Expansion & OA             & Kappa          & F1              \\ \hline
\multirow{4}{*}{River}    & 32         & 2         & 97.27          & 82.09          & 83.57          \\
                          & 32         & 3         & 96.89          & 80.73          & 82.41          \\
                          & 16         & 2         & 97.06          & 81.22          & 82.82          \\
                          & 16         & 3         & \textcolor{red}{97.46} & \textcolor{red}{83.15} & \textcolor{red}{84.53} \\ \hline
\multirow{4}{*}{Farmland} & 32         & 2         & 98.33          & 95.83          & 96.99          \\
                          & 32         & 3         & 98.53          & 96.35          & 97.37          \\
                          & 16         & 2         & 98.44          & 96.12          & 97.20           \\
                          & 16         & 3         & \textcolor{red}{98.58} & \textcolor{red}{96.47} & \textcolor{red}{97.46} \\ \hline
\multirow{4}{*}{USA}      & 32         & 2         & 96.38          & 88.92          & 91.2           \\
                          & 32         & 3         & 96.40          & 89.00          & 91.26          \\
                          & 16         & 2         & 96.44          & 89.17          & 91.41          \\
                          & 16         & 3         & \textcolor{red}{96.95} & \textcolor{red}{90.81} & \textcolor{red}{92.74} \\ \hline
\end{tabular}}
\label{table4}
\end{table}

To analyze the influence of the number of selected bands and channels in feature map on change detection results, we conducted experiments with varying down-sample rates $r_{d}$ and channel expansion rates $r_{e}$. Specifically, we selected $1/r_{d}$ bands from all bands and set the number of channels in feature map to a factor of $r_{e}$ times the number of selected bands. The results are shown in Table \ref{table4}.

As we can see, the model performance suffers a more or less decline when the number of selected bands is less. Especially in River and USA datasets, there is an obvious declining trend, which reveals that 1/32 total bands are not enough to provide sufficient change information to achieve excellent change detection. What is more, a smaller expansion rate typically leads a worse performance. This may  because the reduction in parameters caused a decrease in the network's ability to express complex features. We decide to set $r_{d}$ to 16 and $r_{e}$ to 3, because our model achieves the best performance in all datasets with this pair of hyperparameters. While these values may not be optimal, they provide generally good results across three datasets.

\begin{table}[t]
\centering
\caption{Ablation analysis in three datasets.}
\resizebox{\linewidth}{!}{
\begin{tabular}{cccccc}
\hline
\multicolumn{1}{c|}{Dataset}                   & Band Selection & \multicolumn{1}{c|}{Spatial Attention} & OA                 & Kappa              & F1                              \\ \hline
\multicolumn{1}{c|}{\multirow{5}{*}{River}}    & \ding{55}      & \multicolumn{1}{c|}{\ding{55}}                  & 95.04  & 71.23  & 73.88    \\
\multicolumn{1}{c|}{}                          & $\checkmark$   & \multicolumn{1}{c|}{\ding{55}}                  & 96.28  & 76.91  & 78.93 \\
\multicolumn{1}{c|}{}                          & $\checkmark$   & \multicolumn{1}{c|}{typical}                      & 96.85  & 80.00  & 81.71  \\
\multicolumn{1}{c|}{}                          & $\checkmark$   & \multicolumn{1}{c|}{band-specific}               & \textcolor{red}{97.46}  & \textcolor{red}{83.15}  & \textcolor{red}{84.53}  \\ \hline
\multicolumn{1}{c|}{\multirow{5}{*}{Farmland}} & \ding{55}      & \multicolumn{1}{c|}{\ding{55}}                  & 97.97  & 95.43  & 96.96    \\
\multicolumn{1}{c|}{}                          & $\checkmark$   & \multicolumn{1}{c|}{\ding{55}}                  & 98.33  & 95.84  & 97.00   \\
\multicolumn{1}{c|}{}                          & $\checkmark$   & \multicolumn{1}{c|}{typical}                      & 98.40  & 96.03  & 97.14    \\
\multicolumn{1}{c|}{}                          & $\checkmark$   & \multicolumn{1}{c|}{band-specific}               & \textcolor{red}{98.58}  & \textcolor{red}{96.47}   & \textcolor{red}{97.46}   \\ \hline
\multicolumn{1}{c|}{\multirow{5}{*}{USA}}      & \ding{55}      & \multicolumn{1}{c|}{\ding{55}}                  & 96.13  & 88.46  & 90.92   \\
\multicolumn{1}{c|}{}                          & $\checkmark$   & \multicolumn{1}{c|}{\ding{55}}                  & 96.50  & 89.51  & 91.73   \\
\multicolumn{1}{c|}{}                          & $\checkmark$   & \multicolumn{1}{c|}{typical}                      & 96.48  & 89.19  & 91.39   \\
\multicolumn{1}{c|}{}                          & $\checkmark$   & \multicolumn{1}{c|}{band-specific}               & \textcolor{red}{96.95}  & \textcolor{red}{90.81}  & \textcolor{red}{92.74}   \\ \hline
\end{tabular}}
\label{table5}
\end{table}

\subsection{Ablation Experiment}

We conducted an ablation study to evaluate the efficacy of the deep learning based band selection module and the band-specific spatial attention. Table \ref{table5} presents the change detection performance of models with and without these two modules. The baseline is composed of two residual blocks possessing 32 convolutional kernels, and a multi-scale fusion classifier. The results indicate that adding the band selection module to the baseline leads to better performance in the three datasets, demonstrating that a limited number of discriminative bands are sufficient to train a change detection network. We can observe that the band-specific spatial attention also leads to further improved performance. Furthermore, we compare the performance of band-specific spatial attention and typical spatial attention \cite{woo2018cbam}. As we can see, band-specific spatial attention is more effective than naive spatial attention when used with the band selection module. After band selection, the feature distribution differences among bands are significant. Typical spatial attention ignores this point and produces a uniform attention map for all bands, overemphasizing the features of some positions and ignoring the features of other positions. In contrast, band-specific spatial attention generates a customized attention map to each band, individually enhancing the feature discrimination of each band.

\section{Conclusion}

In this study, an end-to-end hyperspectral image change detection network with band selection (ECDBS) is proposed. The proposed network mainly comprises a deep learning based band selection module and cascading band-specific spatial attention (BSA) blocks. The deep learning based band selection can be jointly optimized and end-to-end inference with the subsequent CD models. It can learn to select the bands that are favorable to change detection, improving HSI-CD performance. The BSA block can extract feature of each band using a strategy tailored to its feature distribution. It can sufficiently capture the change information of each band, thereupon improving the discrimination of change feature. We conduct comprehensive experiments on three benchmark datasets compared with eight cutting-edge HSI-CD algorithms, demonstrating the superiority of the proposed ECDBS.

\bibliographystyle{IEEEtran}
\bibliography{ref}

\end{document}